\documentstyle[12pt]{article}
\catcode`\@=11
\@addtoreset{equation}{section}
\def\@biblabel#1{$^{\mbox{tiny\rm #1}}$}
\catcode`@=12
\renewcommand\theequation{\thesection\arabic{equation}}
\newcommand{\bm}[1]{\mbox{\boldmath $#1$}}

\begin{document}

\begin{titlepage}
\vskip 2em
{\LARGE Eddy diffusivities in scalar transport \par}
\vskip 1.5em
\begin{tabular}{l}
 L. Biferale \\
 {\scriptsize\it Observatoire de Nice, BP 229,
		 06304 Nice Cedex 4, France.}  \\
                     \\
 A. Crisanti \\
 {\scriptsize\it Dipartimento di Fisica,
            Universit\'a di Roma ``La Sapienza'',
            P.le A. Moro 2, I-00185 Roma, Italy.} \\
                     \\
 M. Vergassola \\
 {\scriptsize\it Princeton University, Fluid Dynamics Research Center,
		Princeton, NJ 08544, USA.} \\
                     \\
 A. Vulpiani \\
 {\scriptsize\it Dipartimento di Fisica,
            Universit\'a di Roma ``La Sapienza'',
            P.le A. Moro 2, I-00185 Roma, Italy.} \\
\end{tabular}
\par
\vskip 3em

\centerline{December 20, 1994}
\vskip 3em

\begin{quotation}
Standard and anomalous transport in incompressible flow is
investigated using multiscale techniques.
Eddy-diffusivities emerge from the
multiscale analysis through the solution of an auxiliary equation.
{}From the latter it is derived an upper bound to eddy-diffusivities,
valid for both static and time-dependent flow.
The auxiliary problem is
solved by a perturbative expansion in powers
of the P\'eclet number resummed by Pad\'e approximants and by
a conjugate gradient method.
The results are compared to numerical simulations of
tracers dispersion for three flows having different properties of
Lagrangian chaos. It is shown on a concrete example
how  the presence of anomalous
diffusion can be revealed from the singular behaviour of the eddy-diffusivity
at very small molecular diffusivities.
\vspace{\bigskipamount}

PACS number: 47.27.Qb

Submitted to: Physics of Fluids.

\end{quotation}
\end{titlepage}
\newpage

\section{Introduction }
\label{sec:intro}

The problem of passive scalars diffusion in incompressible velocity
fields has a  theoretical and practical importance in many
fields of science and engineering, ranging from mass and heat transport in
geophysical flows to chemical engineering and combustion \cite{M83}.
The main interest is in the
understanding of the  mechanisms  leading to transport enhancement.
Taking into account the molecular diffusion, the motion of a fluid element
can be described by the following Langevin equation
\begin{equation}
{{d {\bm x}} \over {d t}}= {\bm v}({\bm x}, t) + {\bm \eta}(t),
   \label{eq:1.1}
\end{equation}
where ${\bm v}({\bm x}, t)$ is the Eulerian incompressible
velocity field at the
 position ${\bm x}$
and time $t$, and ${\bm \eta}$ is a Gaussian
white noise with zero mean and correlation function
\begin{equation}
\langle\eta_{i}(t)\, \eta_{j}(t')\rangle= 2 \, D_0\,\delta_{ij}\,
     \delta(t-t').
    \label{eq:1.2}
\end{equation}
The coefficient $D_0$ is the (bare) molecular diffusivity.
If $\theta( {\bm x},t)$
denotes the concentration of tracers, the Fokker-Planck equation \cite{C43}
associated to (\ref{eq:1.1}) is
\begin{equation}
\partial_t\theta + ({\bm v} \cdot{\bm \partial})\,\theta=
                        D_0 \,\partial^2\theta .
    \label{eq:1.3}
\end{equation}
The incompressibility condition
${\bm \partial}\cdot {\bm v}= 0$ is explicitly used in
(\ref{eq:1.3}).
Our interest will be mainly concentrated on
the long-time behaviour of (\ref{eq:1.3}).
For time scales much
longer than the characteristic microscopic time,  the
evolution of $\theta({\bm x},t)$
is dominated by long-wave disturbances.
The equation for these slow modes can be obtained by the usual
``hydrodynamic'' analysis \cite{M890}
\begin{equation}
\partial_t \langle\theta\rangle= D^{E}_{ij}\,
             {\partial^2\over\partial x_i\, \partial x_j}\,
               \langle\theta\rangle
             + \ldots \qquad\qquad
             i,j=1,\ldots,d
    \label{eq:1.4}
\end{equation}
where $\langle\theta\rangle$ is the concentration field
averaged locally over a
volume of linear dimensions much larger than the typical length $l$ of the
velocity field and $d$ is the space dimension.
The corrections in (\ref{eq:1.4})
involve terms containing at least three derivatives of $\langle\theta\rangle$,
which can be neglected in the weak gradients limit
$|{\bm \partial} \langle\theta\rangle|/\langle\theta\rangle\ll l^{-1}$.
Eq. (\ref{eq:1.4}) then reduces to a diffusion equation,
with an effective diffusion tensor $D^{E}_{ij}$ (the eddy-diffusivity tensor).
The latter has a direct practical importance
since it measures the spreading for very long times of a spot of
tracers\,:
\begin{equation}
D^{E}_{ij}= \lim_{t\to\infty}\, {1\over 2t}\,
               \left\langle\bigl(x_i(t)-\langle x_i\rangle\bigr)\,
                           \bigl(x_j(t)-\langle x_j\rangle\bigr)
                \right\rangle\, ,
                         \qquad\qquad
             i,j= 1,\ldots,d.    \label{eq:1.5}
\end{equation}
where ${\bm x}(t)$ is the position of a tracer at time $t$
and the average is
taken over the initial positions or, equivalently, over an ensemble of test
particles. Note that the existence of the limit in
(\ref{eq:1.5}) ensures that the transport is a standard diffusion process,
at least for a very large time. This is the typical
situation, but there are also cases showing the so-called
anomalous diffusion\,: the spreading of particles does not
vary linearly with time but as a power law $t^{\gamma}$, with
$\gamma\neq 1$ (where $\gamma > 1$ and  $\gamma< 1$
correspond to superdiffusive and subdiffusive behaviours, respectively).
Transport anomalies indicate the presence of strong correlations
in the dynamics, even at large time and space scales.
An interesting possibility is the one discussed in \cite{ZAS}.
The flow is periodic, but the Lagrangian phase space is a complicated
self-similar structure of islands and cantori. Particles are thus transported
in a coherent way longer and longer as $D_0$ is deacreased,
finally leading to anomalous diffusion in the absence of any
molecular diffusion.

The aim of this paper is using multiscale techniques \cite{BLP}
to study standard and anomalous diffusion. The multiscale formalism is
introduced in Section~II, where the calculation of eddy-diffusivities
is reduced to the solution of an auxiliary
equation. From the latter, upper and lower bounds for
eddy-diffusivities are derived in the general case of time-dependent flows.
The calculation of the exact analytical expression
of the eddy-diffusivity for parallel flows and
random flows $\delta$-correlated in time is also reviewed.
Random flows with a short correlation time are discussed more
thoroughly in Appendix~1. Numerical methods
are generally needed to solve the auxiliary equation leading
to the eddy-diffusivity tensor for a generic flow.
Two possibilities are discussed in Section~III. The first one
is to perform a Pad\'e resummation of the series expressing
the eddy-diffusivity in powers of the P\'eclet number.
The second is the use of a conjugate gradient algorithm.
Both methods are used in Section~IV to analyze three flows having a
standard diffusive transport (the $ABC$ \cite{Ar65,He66,Doet86},
the $BC$ and a two dimensional time-dependent flow). The results are
compared to numerical simulations of tracers dispersion, i.e.
numerical integrations of (\ref{eq:1.1}).
The flows have been chosen since they can be considered as prototypes
for three very different situations with respect to
Lagrangian chaos \cite{A84,CFPV91}, i.e. the chaotic properties of
the deterministic equation obtained from (\ref{eq:1.1}) suppressing
the noise. If the latter equation is integrable, as for the $BC$ flow,
the diffusion process is
expected to be strongly sensitive to the detailed geometric structure
of the Eulerian field and to the presence of molecular diffusion.
For a non-integrable flow we expect a competition
between coherent transport in the non-chaotic regions
(which turns out to be dominant in the $ABC$ flow) and random advection.
The limiting case is the one of a strongly turbulent flow, like the
time-dependent flow,
where molecular transport can be ignored on a large range of
scales and chaotic advection is dominant.
The study of anomalous diffusion is presented in Section~V,
where the flow introduced in \cite{ZAS} is analyzed.
A singular behaviour of the eddy-diffusivity at high P\'eclet numbers is shown
to be a signature of anomalous transport. A
reliable procedure for predicting the presence of anomalous diffusion
is thus provided.

\section{The multiscale technique}
\label{sec:mst}

A general method for studying transport processes is the so-called
multiscale technique (also known as homogenization \cite{BLP}).
The idea is to exploit the scale separation in the dynamics.
Specifically, let ${\bm v}({\bm x},t)$ be an incompressible velocity field,
periodic both in space and time. (The technique can be extended
to handle the case of a random, homogeneous and stationary velocity field
with some non-trivial modifications in the rigorous
proofs of convergence \cite{AM}.) The scalar field $\theta({\bm x},t)$
evolves according to the Fokker-Planck equation (\ref{eq:1.3}).
The units are chosen in such a way that the
periodicities of ${\bm v}$ are $O(1)$.
In order to avoid trivial sweeping effects,
the average of
the velocity field over the periodicities is supposed to vanish.
We shall be interested in the dynamics of the field $\theta$ on
{\it large} scales assumed to be $O(1/\epsilon)$,
where $\epsilon\ll 1$ is the parameter controlling the
scale separation. Because we expect the scalar field to have a diffusive
dynamics, the associated time scale is $O(1/{\epsilon}^2)$.

The presence of the small parameter $\epsilon$
naturally suggests to look for a perturbative
approach. The perturbation is however singular \cite{BO}
since a constant field is a trivial solution
of (\ref{eq:1.3}).
The origin of this phenomenon can be grasped in
the following simple situation. Let the
large-scale field have a single
wavenumber $\epsilon$.
Because of the advection term in (\ref{eq:1.3}),
a small-scale field $\tilde\theta$ is produced and
the wavenumbers
spaced from those of ${\bm v}$ by multiples of $\epsilon$ are
generally excited.
The interaction between the latter modes and those of
${\bm v}$, due again to the advection term,
is responsible for the transport coefficients renormalization.
The essential shift of order
$\epsilon$ in the wavenumbers of $\tilde\theta$ with respect to those
of ${\bm v}$ is missed by regular
perturbation expansions.
Asymptotic methods, like multiscale techniques, are
thus needed.

In addition to the {\em fast}
variables ${\bm x}$ and $t$, let us then introduce {\em slow}
variables as ${\bm X}=\epsilon {\bm x}$ and $T={\epsilon}^{2} t$. The
prescription of the technique is to treat the two sets of variables
as independent. It follows that
\begin{equation}
\label{derivate}
\partial_i\mapsto\partial_i+\epsilon\nabla_i\,;\qquad
\partial_t\mapsto\partial_t+\epsilon^2\partial_T,
\end{equation}
where $\partial$ and $\nabla$ denote the derivatives with respect to
fast and slow space variables, respectively. The solution is sought as a
perturbative series
\begin{equation}
\label{serie}
\theta({\bm x},t;{\bm X},T)=\theta^{(0)}+\epsilon\theta^{(1)}+
\epsilon^2\theta^{(2)}+\ldots ,
\end{equation}
where the functions $\theta^{(n)}$ depend {\it a priori} on both fast
and slow variables. By inserting (\ref{serie}) and (\ref{derivate})
into (\ref{eq:1.3}) and equating terms having equal powers in $\epsilon$, we
obtain a hierarchy of equations. The solutions of interest to us are those
having the same periodicities as the velocity field.
The first equation, corresponding to $O(\epsilon^0)$, is
\begin{equation}
\label{prima}
\partial_t\theta^{(0)}+\left({\bm v}\cdot{\bm \partial}\right)
\theta^{(0)} = D_0\, \partial^2 \theta^{(0)}.
\end{equation}
By using Poincar\'e inequality, one can show \cite{UF} that for periodic
solutions
\begin{equation}
\label{poincare}
-\partial_t\int\,\left(\theta^{(0)}\right)^2\,dV= D_0 \int
\,\left({\bm \partial}\theta^{(0)}\right)^2\,dV\geq
D_0\,\left({2\pi\over  L}\right)^2\int
\,\left(\theta^{(0)}\right)^2\,dV,
\end{equation}
where $L$ is the spatial periodicity length of ${\bm v}$ (supposed
for simplicity to be the same in all directions) and the
integral is over the periodicity box.
The inequality (\ref{poincare})
implies that the solution will relax to a constant
with respect to fast variables, i.e.
\begin{equation}
\label{sol1}
\theta^{(0)}({\bm x},t;{\bm X},T)=\theta^{(0)}({\bm X},T).
\end{equation}
It can be also easily checked that the
transient has no effect on the large-scale dynamics.
The equations at order $\epsilon$ and $\epsilon^2$ are
\begin{equation}
\label{seconda}
\partial_t\theta^{(1)}+\left({\bm v}\cdot{\bm \partial}\right)
\theta^{(1)} - D_0\, \partial^2 \theta^{(1)} = -{\bm v}\cdot\nabla
\theta^{(0)},
\end{equation}
\begin{equation}
\label{terza}
\partial_t\theta^{(2)}+\left({\bm v}\cdot{\bm \partial}\right)
\theta^{(2)} - D_0\, \partial^2 \theta^{(2)} =
-\partial_T\theta^{(0)}-({\bm v}\cdot\nabla )
\theta^{(1)} + D_0\, \nabla^2 \theta^{(0)}
+2D_0{\bm \partial}\cdot\nabla\theta^{(1)}.
\end{equation}
Since the equation (\ref{seconda}) is linear,
its solution can be written as
\begin{equation}
\label{sol2}
\theta^{(1)}({\bm x},t;{\bm X},T)=\theta^{(1)}({\bm X},T)
+{\bm w}({\bm x},t)\cdot\nabla\theta^{(0)}({\bm X},T),
\end{equation}
where the first term on the r.h.s. is a solution of the homogeneous equation
and the vector field ${\bm w}$ has a vanishing average
over the periodicities and satisfies
\begin{equation}
\label{chi}
\partial_t{\bm w}+\left({\bm v}\cdot{\bm \partial}\right)
{\bm w} - D_0\, \partial^2{\bm w} = -{\bm v} .
\end{equation}
Due to the incompressibility of the velocity field,
the average over the periodicities of the
l.h.s. in (\ref{seconda}) and (\ref{terza})
is zero. For the equations to
have a solution, the average of the r.h.s. should also
vanish (Fredholm alternative). The resulting solvability conditions
provide the equations governing the large-scale dynamics, i.e. the dynamics in
the slow variables. From (\ref{terza}) we obtain
\begin{equation}
\label{solv}
\partial_T\langle\theta^{(0)}\rangle=D_0\, \nabla^2\langle\theta^{(0)}\rangle
-\langle{\bm v}\cdot\nabla \theta^{(1)}\rangle,
\end{equation}
where the symbol $\langle\cdot\rangle$ denotes the average over the
periodicities.
The solvability condition for (\ref{seconda}) is trivially satisfied,
reflecting the absence of $\alpha$-type effects \cite{HKM,FSS}.
By plugging (\ref{sol2}) into (\ref{solv}) we obtain the diffusion equation
\begin{equation}
\label{diffusion}
\partial_T\theta^{(0)}({\bm X},T)
=D^{E}_{ij}\,\, \nabla^2 \theta^{(0)}({\bm X},T),
\end{equation}
where the eddy diffusivity tensor is
\begin{equation}
\label{eddydiff}
D^{E}_{ij}=D_0\delta_{ij}-{1\over 2}\left[\,\langle v_i w_j\rangle+
\langle v_j w_i\rangle\,\right].
\end{equation}
Remark that the structure of
the eddy-diffusivity tensor will reflect
the rotational symmetries of ${\bm v}$ and is in general non-isotropic.

\subsection{Inequalities for the eddy-diffusivity}

Two important inequalities
can be derived from the auxiliary equation (\ref{chi})
and the expression (\ref{eddydiff})
of the eddy-diffusivity.
Let us consider the $i$-th and the $j$-th
components of (\ref{chi}) and multiply by $w_j$ and $w_i$, respectively.
Taking the sum and averaging, the time derivative
and the advective term vanish and we obtain
\begin{equation}
\label{clue}
-{1\over 2}\left[\,\langle v_i w_j\rangle+
\langle v_j w_i\rangle\,\right]=
        D_0\langle{\bm \partial}w_i\cdot {\bm \partial}w_j\rangle.
\end{equation}
{}From (\ref{clue}) and (\ref{eddydiff}) it follows
\begin{equation}
\label{positivo}
D^{E}_{ij}=D_0\left[
\delta_{ij}+ \langle{\bm \partial}w_i\cdot {\bm \partial}w_j\rangle\right].
\end{equation}
This expression of the eddy-diffusivity clearly shows that the
correction to the molecular contribution is positive definite.
Large-scale scalar transport is therefore enhanced in the presence
of a small-scale incompressible velocity field.
The cause is that for the advection-diffusion equation
(\ref{eq:1.3}) the integral of $\theta^2$ over the whole space
is a decreasing function of time.
When the dynamics does not possess the latter property the large-scale
transport can actually be depleted, rather than increased.
For momentum transport
in Navier-Stokes flow the depletion can be so strong that the eddy-viscosity
becomes negative, i.e. the average flux is
in the same direction as the large-scale gradient \cite{SY,GVF}.

The second inequality also is derived from (\ref{clue}) but it
is an upper bound to
eddy-diffusivities. Because of incompressibility, the velocity field
can be expressed using a vector potential as
${\bm v}={\rm rot}\,{\bm A}$.
By taking the trace of (\ref{clue}) and integrating
by parts, we obtain
\begin{equation}
\label{start}
0\leq   D_0 \langle{\bm \partial}w_i\cdot {\bm \partial}w_i\rangle=
 -\langle v_i w_i\rangle=
 -\langle{\bm A}\cdot\,{\rm rot}\,{\bm w}\rangle.
\end{equation}
Application of the Schwartz inequality leads to
\begin{equation}
\label{schwartz}
D_0 \langle{\bm \partial}w_i\cdot {\bm \partial}w_i\rangle \;\;\;\leq \;\;\;
\langle A^2\rangle^{1/2}\langle\left({\rm rot}\,{\bm w}\right)^2\rangle^{1/2}
\;\;\;
\leq  \;\;\; \langle A^2\rangle^{1/2}
\langle {\bm \partial}w_i\cdot {\bm \partial}w_i\rangle^{1/2},
\end{equation}
whence
\begin{equation}
\label{lthan}
{D^{E}_{ii}\over D_0}\leq\; d+{\langle A^2\rangle\over D_0^2}\equiv
d+{\rm Pe}^2.
\end{equation}
The P\'eclet number is denoted by ${\rm Pe}$.
The result (\ref{lthan}), valid for time-dependent flows also,
generalizes a similar inequality known for time-independent velocity
fields \cite{AM,KST}.
The inequality (\ref{lthan}) also provides an upper bound for each eigenvalue
since the eddy-diffusivity tensor is positive definite.

\subsection{Two exactly solvable cases}

By using multiscale techniques,
the calculation of eddy diffusivities has been reduced
to the solution of the auxiliary equation (\ref{chi}).
Numerical methods are generally needed to solve it
but there are a few cases where one can obtain the solution of
(\ref{chi}) analytically.
We shall briefly review here the case of parallel flows and
random flows $\delta$-correlated in time.

The peculiar property of
parallel flows is that the velocity is everywhere
in the same direction, e.g. in three dimensions
\begin{equation}
\label{parallel}
{\bm v}(x,y,z;t)= \left( v_x(y,z;t),0,0\right),
\end{equation}
and $v_{x}$ cannot depend on $x$ because of incompressibility.
The advective non-linearity ${\bm v}\cdot\partial\,{\bm v}$ is thus
vanishing. Thanks to the latter,
we can easily obtain the solution of the auxiliary equation (\ref{chi})
as
\begin{equation}
\label{risolvi}
\hat{w}({\bm q},\omega)= {\hat{v}({\bm q},\omega)\over
i\omega - D_0 q^2}.
\end{equation}
The Fourier transforms of $v_{x}$ and $w_{x}$ are denoted by
$\hat{v}$ and $\hat{w}$.
If $F({\bm q},\omega)=<|\hat{v}({\bm q},\omega)|^2>$,
it follows that the eddy-diffusivity is
\begin{equation}
\label{risolto}
D^{E}_{\parallel}=D_0\,\left(1+\int {F({\bm q},\omega)\,q^2\over
\omega^2+D_0^2q^4}\,d{\bm q}\,d\omega\right)\; ;\qquad
D^{E}_{\perp}=D_0 .
\end{equation}
Here, $D^{E}_{\parallel}$ and $D^{E}_{\perp}$
are the components of the eddy-diffusivity tensor
parallel and orthogonal to the direction of the velocity.
For the Kolmogorov flow $v_{x}=V\cos\,y$ and the
parallel eddy-diffusivity $D^{E}_{\parallel}=D_0 + V^{2}/2D_0$.

Let us now consider random flows having a short correlation time $\tau$.
Neglecting the diffusion term in (\ref{chi})
we obtain a hyperbolic equation which can be formally integrated
along the characteristics
\begin{equation}
\label{hyper}
{\bm w}\left({\bm x}({\bm a},t);t\right)= -\int_0^t
{\bm v}\left({\bm x}({\bm a},s);s\right)\,ds +{\bm w}({\bm a};0).
\end{equation}
Here, ${\bm a}$ denotes the Lagrangian initial position and the Eulerian
position at time $t$ is
\begin{equation}
\label{position}
{\bm x}({\bm a},t)={\bm a}+\int_0^t{\bm v}
\left({\bm x}({\bm a},s);s\right)\,ds .
\end{equation}
{}From (\ref{hyper}) Taylor's expression of the eddy-diffusivity tensor
immediately follows
\begin{equation}
\label{taylor}
D^{E}_{ij}={1\over 2}\int_0^{\infty}
\left(\Gamma^L_{ij}(s)+\Gamma^L_{ji}(s)\right)\,ds ,
\end{equation}
where the Lagrangian correlation function is defined as
\begin{equation}
\label{lagrange}
\Gamma^L_{ij}(t-s)= \langle v_i \left({\bm x}({\bm a},t);t\right)
v_j\left({\bm x}({\bm a},s);s\right)\rangle.
\end{equation}
The operation $\langle\cdot\rangle$ denotes either
spatial or ensemble averaging which do coincide
since the velocity field is supposed homogeneous, stationary
and mixing. Note that the convergence of the integral (\ref{taylor})
is not at all guaranteed. The role of a small, but
non-zero, molecular diffusivity can be crucial in this respect \cite{M83}.
In the limit where $\tau$ is small, the Lagrangian correlation
$\Gamma^{L}_{ij}$ tends to the Eulerian correlation $\langle v_{i}({\bm x},t)
\,v_{j}({\bm x},s)\rangle$. For a signal $\delta$-correlated in time
\begin{equation}
\label{delta}
\langle v_i({\bm x},t)\,v_j({\bm x},s)\rangle=2\,F_{ij}\delta(t-s),
\end{equation}
and the expression (\ref{taylor}) reduces to
\begin{equation}
\label{eccoci}
D^{E}_{ij}=D_0\,\delta_{ij}+F_{ij}.
\end{equation}
The corrections to this result
due to a small, but finite, correlation time will
be studied in Appendix~1.

\section {Numerical methods}
\label{sec:numsim}

Whenever the auxiliary equation (\ref{chi}) cannot be solved exactly,
numerical methods are needed. In this Section we shall discuss two different
methods that
we have used\,: a perturbative expansion and a conjugate gradient
algorithm.

\vskip 0.2cm
In the perturbative method, the solution ${\bf w}$ of the auxiliary
equation (\ref{chi}) is sought as a power series in the
P\'eclet number ${\rm Pe} \sim 1/D_0$:
\begin{equation}
 {\bm w} = {\rm Pe}\,{\bm w}^{(1)}
                         + {\rm Pe}^2\,{\bm w}^{(2)} + \cdots.
\label{eq:series}
\end{equation}
We shall concentrate on the time-independent case for simplicity.
By inserting the expansion (\ref{eq:series}) into (\ref{chi}) the following
recursive relation is obtained
\begin{equation}
\label{eq:hierar}
{\bm w}^{(1)} = \partial^{-2}{{\bm v}\over D_0\,{\rm Pe}},\quad
{\bm w}^{(2)} = \partial^{-2}{{\bm v}\cdot\partial\,{\bm w}
^{(1)}\over D_0\,{\rm Pe}},\quad\ldots\quad
{\bm w}^{(n)} = \partial^{-2}
{{\bm v}\cdot\partial\,{\bm w}^{(n-1)}\over D_0\,{\rm Pe}},\,\ldots
\end{equation}
Expressions (\ref{eq:hierar}) and the calculation of the
average value in (\ref{eddydiff}) are conveniently handled in
Fourier space, leading to
\begin{equation}
\label{eq:dseries}
{D^E_{ij}\over D_0} = \delta_{ij} + \sum_{n\geq 1} (c_n)_{ij}
\,{\rm Pe}^{2n}.
\end{equation}
Here, the $c_n$'s are numerical coefficients
and the series turns out to be in ${\rm Pe}^2$, rather than in
${\rm Pe}$.
The contribution of order
$2n+1$ in $<v_i\,w_j>$ is indeed antisymmetric, as can be easily checked
by integrating $n$ times by parts. The series (\ref{eq:dseries})
will in general converge for ${\rm Pe}<{\rm Pe}^*$ only,
because of singularities in the complex plane.
A reliable analytic continuation  beyond the disc of convergence
can however be performed. In \cite{AMX} it was indeed shown
that the component of the
eddy-diffusivity in the arbitrary direction $\hat{n}$
can be represented as a Stieltjes integral
\begin{equation}
\label{cacchio}
{D^{E}_{\hat{n}}\over D_0}=1+{\rm Pe}^2\int dz
{{\rho}_{\hat{n}}(z)\over
1+{\rm Pe}^2 z^2},
\end{equation}
where $\rho_{\hat{n}}(z)$ is a positive definite function, possibly singular.
The poles of the eddy-diffusivity,
considered as a function of a complex variable, are
all on the imaginary axis. Moreover, it follows from (\ref{cacchio}) that
Pad\'e approximants
of (\ref{eq:dseries}) have some interesting peculiar
properties (see e.g. \cite{BO}).
Let us indeed denote by $P_n^n({\rm Pe})$ the diagonal Pad\'e approximant of
order $n$ for the series (\ref{eq:dseries})
and by $P_{n+1}^n({\rm Pe})$
the Pad\'e approximant having the
numerator and the denominator of degree $n$ and $n+1$, respectively.
The following results hold for every value of the P\'eclet number\,:
(i) The diagonal sequence $P_n^n$ is monotonically increasing
and has an upper bound\,; (ii) The sequence $P_{n+1}^n$ is monotonically
decreasing and has a lower bound\,;
(iii) The exact value $P^{\star}$ of the Stieltjes integral
satisfies
\begin{equation}
\label{soprasotto}
\lim_{n\to\infty}P_{n}^n\leq P^{\star}\leq
\lim_{n\to\infty}P_{n+1}^n.
\end{equation}
The difference $\left( P_{n+1}^n-P_{n}^n\right)$
decreases monotonically in $n$ and provides an upper bound
to the error due to the finite order.
The quality of the resummation by a finite order
approximant can be thus checked self-consistently.
Pad\'e approximants are very sensitive
to the precision in the computations when the series is extended
well beyond its radius of convergence. For small values of the
molecular diffusivity, the coefficients in the series (\ref{eq:dseries})
must be then known with very high precision.
In our numerical calculations we used the
FORTRAN multiple-precision package MP, written by
R.P. Brent \cite{Br81}. It should be noted, however,
that very high precision computations are quite expensive
in computers memory costs (see next Section).

\vskip 0.2cm
The second method that we have used to solve the auxiliary equation
(\ref{chi}) is a conjugate gradient algorithm \cite{cgradient}.
The components of the vector ${\bf w}$
are not coupled in eq. (\ref{chi}), which is thus equivalent to a set
of scalar equations. All of them
can be written in Fourier space as
\begin{equation}
A_{i,j}\,x_j=b_i \qquad i,j=1,\ldots ,V.
\label{eq:A.6}
\end{equation}
Here, $V$ is the resolution,
$x_i$ and $b_i$ are vectors having the $V$ components
equal to the Fourier transform of the relevant components of
${\bm w}$ and
$-{\bm v}$, respectively. Conjugate gradient algorithms are
widely used to minimize multidimensional
functions when the number of dimensions is very large.
The interested reader is referred to \cite{rossi}
for a comparison with other methods
(Gauss-Siedel or Minimal-Residue \cite{cgradient})
in another stiff numerical problem, the inversion
of the propagator in lattice quantum chromodynamics.
The solution of the problem (\ref{eq:A.6}) is sought by minimizing
the quantity $(Ax-b)^2$ over a sequence
of directions orthogonal  to
the matrix $A$. In all the applications
of the method that we have considered, the matrix $A$ in (\ref{eq:A.6}) is
sparse (quasi-diagonal). Each iteration of the minimization algorithm
can be then performed in $O(V)$ operations, rather than  $O(V^2)$.
For a positive-definite matrix, the rate of convergence of the method
can be shown to be exponential \cite{cgradient}.
Our matrix $A$ has actually one zero eigenvalue,
corresponding to a constant field.
The problem is nevertheless well-posed
since both the velocity field and the solution ${\bm w}$ are orthogonal
to constants, i.e. have zero average.
As in any other numerical scheme, the simulations are expected to
become more and more demanding as the molecular diffusivity becomes
smaller. An increasing number of excited scales requires indeed a greater
resolution and the rate of convergence of the method decreases when
$V$ and the P\'eclet number are increased. It is also to be checked
that no eigenvalue is equal to zero within the numerical accuracy
because of round-off errors.
In the next Section it will turn out that
the previous limitations are not very severe
and do not forbid to perform high P\'eclet numbers simulations.

\vskip 0.2cm
We conclude this Section by briefly describing the numerical scheme
used for the numerical simulations of tracers dispersion.
The latter are done by
uniformly distributing $N$ particles in the periodicity box
and letting them evolve according to the Langevin equation
(\ref{eq:1.1}). The $i$-th diagonal element of the eddy-diffusivity
tensor is then given by
\begin{equation}
  \sigma_i^2(t) =  \lim_{t\to\infty} {1\over 2Nt}
		   \,\,\sum_{k=1}^N \left[
                   x_i^{(k)}(t) - {1\over N}\,\sum_{j=1}^N x_i^{(j)}(t)
                                           \right]^2.
\label{eq:A.1}
\end{equation}
The indices $k$ and $j$ label the $N$ particles whereas the index $i$
denotes the spatial directions ($x,y$ for the two-dimensional
and $x,y,z$ for the three-dimensional case).
The numerical integration of the Langevin
equation was performed by
a Runge-Kutta algorithm, modified to take into account the white noise
term \cite{Ho92}. The integration step was $\Delta t= 0.01$ and
the total number of integration steps was $10^6$. This ensured a good
convergence of the quantities (\ref{eq:A.1}) also for the lowest molecular
diffusion coefficients $D_0$ used. The number of particles used was
$1\,000$ for the three-dimensional and $2\,000$ for the
two-dimensional case.

\section{Standard diffusion}

\label{sec:new}

The aim of this Section is to apply the methods previously discussed
to three flows showing standard diffusion. The criterion in the choice
of the flows is to have different mechanisms of diffusion
enhancement, highlighting the influence of Lagrangian chaos
on transport at high P\'eclet numbers. Specifically, we have considered\,:

\begin{itemize}
\item The three dimensional ABC flow \cite{Ar65,He66,Doet86}\,:
\begin{equation}
  \cases{
         \dot x & $= A \sin (z) + C \cos (y) $, \cr
         \dot y & $= B \sin (x) + A \cos (z) $, \cr
         \dot z & $= C \sin (y) + B \cos (x) $. \cr
        }
\label{eq:A.3}
\end{equation}
with $A=B=C$. The $ABC$ flow is a Beltrami time-independent
solution of Euler's equations.
Eq. (\ref{eq:A.3}) shows Lagrangian chaos but the phase space is also
made of regular regions, having roughly the shape of a tube parallel
to one of the three axes (principal vortices).

\item The two dimensional BC flow
\begin{equation}
  \cases{
        \dot x & $= C \cos (y) $, \cr
        \dot y & $= B \cos (x) $, \cr
        }
\label{eq:A.4}
\end{equation}
obtained by projecting the flow (\ref{eq:A.3}) onto the $x-y$ plane
and translating the $x$-coordinate
by $\pi/2$. Eq. (\ref{eq:A.4}) is integrable
and the streamlines form a closed structure made of four cells in
each periodicity box.

\item The two dimensional time dependent flow
\begin{equation}
  \cases{
        \dot x & $= \cos (y) + \sin (y)\, \cos(t) $\cr
        \dot y & $= \cos (x) + \sin (x)\, \cos(t) $\cr
        }
\label{eq:A.5}
\end{equation}
This flow is not a solution of Euler's equations anymore
but it is the superposition of the flow (\ref{eq:A.4})
with another
flow of the same type oscillating with frequency $\omega=1$.
The motivation for introducing a time dependency
is to destroy all possible ``regular islands'', like the vortices in
(\ref{eq:A.3}).

\end{itemize}

Note that both the flows (\ref{eq:A.3}) and (\ref{eq:A.4}) have
an isotropic eddy-diffusivity tensor. Let us indeed consider the latter
for simplicity and perform the
following two operations\,:
translation by $\pi$ and mirror-inversion with respect to one of the
axes (e.g. $x\mapsto \pi-x$ and $y\mapsto \pi-y$). From the auxiliary eq.
(\ref{chi}) it follows that, under the previous operations, one of the
components of ${\bf w}$ is odd and the other is even, in such a way that
$<v_x\,w_y>=<v_y\,w_x>=0$. The diagonal components are obviously equal
because of the symmetry $x\leftrightarrow y$.
For (\ref{eq:A.3}) the proof is similar, exploiting
the fact that the group of symmetries of the flow is
isomorphic to the
cubic group \cite{Doet86}. The flow (\ref{eq:A.5}) possesses the symmetry
$x\leftrightarrow y$, but it is not mirror-symmetric. The diagonal
components of the eddy-diffusivity will then be equal
but the non-diagonal component does not vanish.
In Langevin simulations the previous symmetry properties are exploited
to reduce the statistical fluctuations
by averaging over the directions.

In Figs.~1, 2 and 3  we present the results for the diagonal component
of the eddy-diffusivity tensor
of (\ref{eq:A.3}), (\ref{eq:A.4})
and (\ref{eq:A.5}), respectively.
The curves in each figure correspond to
numerical simulations of the Langevin equation, the Pad\'e method and
the conjugate gradient
algorithm. To attain the highest P\'eclet number
the order  of Pad\'e approximants
used is  $54$, $115$, $29$ and the number of significant digits
in the computations is
83, 203, 40, respectively. Concerning the conjugate gradient
algorithm, in Fig.~4 it is shown the power spectrum of the auxiliary
field ${\bf w}$ for the flow (\ref{eq:A.5})
at $D_0=0.01$. It can be seen that the
field is resolved enough to ensure the presence of a conspicuous
exponentially decaying tail. The
conjugate gradient algorithm turns out to be
much more efficient at high P\'eclet numbers than the Pad\'e method.
The latter has the advantage of requiring the calculation of the coefficients
of (\ref{eq:dseries}) only\,: once they are computed, the eddy-diffusivities
for all values of $D_0$ such that the method works are available.
On the other hand, the memory costs for high precision arithmetics
are a major drawback and practically restrict the method to moderate
P\'eclet numbers.

{}From the high P\'eclet number behaviour of the eddy-diffusivities
in the figures it is clear that the three flows have a very different dynamics.
The main contribution to diffusion
in the flow (\ref{eq:A.3})
comes from the particles in the vortices, where the transport is almost
ballistic, leading to the observed $1/D_0$ dependence.
Because of the presence of closed
cells, a non-zero molecular diffusivity is needed
to have an effective diffusion in the flow (\ref{eq:A.4}), as
indicated by the $\sqrt{D_0}$ behaviour in Fig.~2.
The transport for small molecular diffusivities
indeed occurs by jumps from one cell to another
due to the white-noise term
in the Langevin equation \cite{Po85,Roet87,Sh87}.
The probability of jumping is controlled
by the width of boundary layers located near the separatrices and gives
the square-root law. The flow (\ref{eq:A.5}) is finally an example
of strong Lagrangian turbulence. The particles
can diffuse even in the absence of molecular diffusion, chaotic advection
is dominant and the eddy-diffusivity attains a finite value indipendent of
the molecular diffusivity. The figures show that for all the flows
considered,
Langevin simulations and numerical solutions of the auxiliary equation
(\ref{chi}) do agree. Moreover, in the latter method
no problem of finite statistics and simulation times must be overcome.
We conclude that multiscale techniques combined with an efficient
numerical scheme for the solution of the auxiliary equation
(e.g. a conjugate gradient algorithm or a pseudo-spectral
code \cite{pseudo}) provide a powerful tool for the calculation of
eddy-diffusivities and transport properties.

\section{Anomalous diffusion}
\label{sec:anomad}

We shall discuss here how the
multiscale formalism presented in the previous Sections
can be used for the
problem of anomalous diffusion.
At a first sight it would seem that multiscale techniques cannot be used
anymore. Consider indeed  a two-dimensional
static parallel flow (\ref{parallel}). If the
power spectrum $F(q)$ defined in Section~II.2 is such that
\begin{equation}
\label{eq:fqsm}
  F(q) \sim q^{\alpha}, \qquad \alpha \le 1,
   \qquad \mbox{\rm for}\ q\ll 1,
\end{equation}
then the integral in
(\ref{risolto}) diverges and the eddy-diffusivity is not defined.
The divergence is actually reflecting the fact that the transport
in the direction of the flow is
superdiffusive \cite{MaMa80,Re89,Zuet90,CV93}, i.e.
\begin{equation}
\label{eq:superd}
  \langle x^2(t)\rangle \sim t^{2\nu}, \qquad \nu > 1/2,
\end{equation}
and it is not a standard diffusion.
The particles are indeed coherently
swept by large-scale modes having wavelengths
comparable (or even larger) to the typical length of the scalar field.
Scale-separation breaks down and multiscale
methods, heavily relying on this assumption, seem to become useless.
Let us however cut out the singular part in
the integral (\ref{risolto}) by defining a
regularized velocity field ${\bm v}_{L}$, such that
\begin{equation}
\label{eq:regsp}
  F_{L}(q) = \left\{\begin{array}{ll}
                        F(q) &\ \mbox{\rm if}\ q > L^{-1} \\
                        0          &\ \mbox{\rm if}\ q < L^{-1} \\
                          \end{array}
                   \right.
\end{equation}
The eddy-diffusivity is now finite and exhibits a dependence
\begin{equation}
\label{eq:dreg}
 D^E_{\parallel}(L) \sim L^{1-\alpha}, \qquad L\gg 1 ,
\end{equation}
on the cut-off length $L$. A standard diffusion is however observed
only for spatial and time lengths larger than $L$ and $t^*\sim L^2/D_0$,
respectively. For $t\sim t^*$ the system has indeed a crossover \cite{Yo91}
and for times shorter than $t^*$ it shows the same behaviour as in
(\ref{eq:superd}). By matching at $t^*$
the two different regimes, we
obtain
\begin{equation}
\label{eq:nu}
  \nu = \frac{3-\alpha}{4} \ge 1/2.
\end{equation}
For $\alpha=0$, i.e. a velocity field which is a white noise in space,
(\ref{eq:nu}) leads to $\nu=3/4$, the well known result of Matheron and
De Marsily \cite{MaMa80}.

In the previous example the origin of superdiffusion was related to
the spatial
structure of the velocity field.
Another interesting case is the one of
velocity fields with very long Lagrangian correlation times.
The integral defining the eddy-diffusivity
in Taylor's expression (\ref{taylor}) may then diverge, indicating
the presence of anomalous transport for $D_0=0$.
Note however that for any $D_0>0$ the transport is
in general a standard diffusion (see (\ref{risolto})
for an example of the role of a small
molecular diffusivity).
The molecular diffusivity can be thus used as a regularization parameter,
similarly to the cut-off length for parallel flows.
As in the latter case, by studying
the behaviour of the eddy-diffusivity close to the critical
point (small $D_0$)
one should be able to have some insights into the anomalous
behaviour at the critical point ($D_0=0$).
Specifically, the examples of Section~IV
show that in the presence of ballistic channels the eddy-diffusivity
varies as the inverse of $D_0$ for small
$D_0$, while for a system with strong Lagrangian chaos it
tends to a constant. If $D^E_{\hat{n}}$ denotes
the eddy-diffusivity in the arbitrary direction $\hat{n}$, we are thus led to
interpret a small $D_0$ behaviour
\begin{equation}
\label{eq:scaling}
 D^E_{\hat{n}}\sim D_0^{-\beta}, \qquad 0<\beta<1
\end{equation}
as a mark of anomalous diffusion in the direction $\hat{n}$ for $D_0=0$.
For a practical application of the previous argument we have considered the
velocity field \cite{ZAS}
\begin{equation}
\label{eq:zas}
 \left\{
   \begin{array}{ll}
     \dot{x} &= \phantom{-}\partial_y \psi + \varepsilon\sin z
       \nonumber \\
     \dot{y} &= -\partial_x \psi+ \varepsilon\cos z
       \nonumber \\
     \dot{z} &= \phantom{-}\psi
    \end{array}
 \right.
\end{equation}
where
\begin{equation}
\label{eq:strezas}
 \psi(x,y) = 2\left[ \cos x + \cos\left(\frac{x+\sqrt{3}\,y}{2}\right)
                            + \cos\left(\frac{x-\sqrt{3}\,y}{2}\right)
              \right] .
\end{equation}
Numerical simulations of (\ref{eq:zas}) have led the authors of
\cite{ZAS} to conclude that the flow exhibits
anomalous diffusion in the $x-y$ plane for some intervals
of $\varepsilon$-values in the range $(0,5)$. In particular,
$\varepsilon=1$ and $\varepsilon=2.3$ are such that the diffusion is standard
and anomalous, respectively. Let us now introduce a small molecular
diffusivity and calculate the eddy-diffusivity of the flow
(\ref{eq:zas}). The auxiliary
equation is solved by the conjugate gradient method and the results for
$D_{xx}$ are presented in Figs.~5 and 6.
It is evident that for $\varepsilon=1$ the eddy-diffusivity
tends to a constant for small $D_0$, while for $\varepsilon=2.3$ it is
observed the
behaviour $D_{xx}\sim D_0^{-\beta}$ with $\beta \simeq 0.7$.
This value is in rough agreement with the numerical results of \cite{ZAS}
using the dimensional relation $\beta = 2\nu -1$ obtained by
matching the diffusive behaviour with the anomalous behaviour (\ref{eq:superd})
at the typical diffusive time $O(1/D_0)$.
The criterion (\ref{eq:scaling}) is thus confirmed and we conjecture that
its validity is not restricted to the flow (\ref{eq:zas}) only. For
a generic flow, anomalies in the zero-diffusivity dynamics
could be then captured by introducing a small
molecular diffusivity and looking for a singular behaviour of
transport coefficients.
As shown in the previous section, the advantage with respect to
simulations of the Langevin equation is that
no problem of statistical fluctuations must be
tackled. The previous procedure should then allow to make robust predictions
on the presence of anomalous transport.

\bigskip
\noindent{\bf Acknowledgments}

\noindent
We are grateful to M.~Avellaneda and U.~Frisch for extensive discussions.
Two of us (AC and AV) acknowledge the
organizers and participants of the workshop
{\it Dynamics of Transport in Fluids, Plasmas and Charged Beams}
(ISI, Torino, July 1994) for stimulating discussions.
We thank the ``Istituto di Cosmogeofisica" of CNR, Torino, were part of
the numerical analysis was done. LB was
supported by the EEC contract ERBCHBICT941034.
The work of MV was supported by ONR/DARPA under Grant
No. N00014-92-J-1796 P00001.

\newpage
\section*{Appendix 1}
\label{s:appendix1}
\renewcommand{\theequation}{A.\arabic{equation}}
\setcounter{equation}{0}

We shall derive here
the expression of the eddy-diffusivity for incompressible
flows having a short correlation time. Specifically,
the ratio $\tau_c/\tau_s$ between
the correlation and the sweeping time (defined precisely later)
is supposed to be small. We shall be particularly interested
in 3D isotropic flows.
By the latter we mean velocity fields invariant under rotations,
but not in general under parity transformations.
When the correlation time tends to zero, we recover the case
of flows $\delta$-correlated in time. For a Gaussian flow,
the first two
corrections are shown to be
proportional to $\left(\tau_c/\tau_s\right)^2$\,:
one of them is related to the
correlation length of the flow and the second is due to helicity.
The former reduces, while the latter
increases the eddy-diffusivity, in agreement with \cite{RHK1}.

Let ${\bm v}({\bm x},t)$
denote a random, homogeneous and stationary
incompressible velocity field. We shall suppose the flow to be isotropic,
Gaussian and the
correlation function
\begin{equation}
\label{spectrum}
\langle v_i({\bm x},t)\,v_j({\bm y},s)\rangle=
C(|t-s|)\,B_{ij}({\bm x}-{\bm y})
\; .
\end{equation}
The mean velocity is equal to zero.
The temporal correlation function $C(t)$ decays
on a time-scale of order $\tau_c$. The spatial correlation function is defined
via its Fourier transform as
\begin{equation}
\label{Fourier}
\hat{B}_{ij}({\bm k})= P_{ij}({\bm k})\,{E(k)\over 4\pi k^2}
- {i\over 2}\epsilon_{ijl}\,k_l\,{H(k)\over 4\pi k^4} \; ,
\end{equation}
where $P_{ij}=\delta_{ij}-k_ik_j/k^2$ is the solenoidal projector and
$\epsilon_{ijk}$ is the fundamental antisymmetric tensor.
The functions $E(k)$ and
$H(k)$ will be called the energy and
the helicity spectrum since
\begin{equation}
\label{energy}
{1\over 2}\langle v^2\rangle=C(0)\int E(k)\,dk\,;\quad
\langle{\bm v}\cdot\omega\rangle=C(0)\int H(k)\,dk\;.
\end{equation}
The helicity is a pseudo-scalar and it is thus vanishing for flows having
a center of symmetry (parity-invariance).
The helicity spectrum satisfies the inequality (see \cite{ML})\,:
\begin{equation}
\label{Keith}
|H(k)|\leq 2\,k\,E(k)\;.
\end{equation}

We shall be interested in the calculation of eddy-diffusivities for very
high P\'eclet numbers. The eddy-diffusivity is
given in this limit by Taylor's expression (\ref{taylor})
\begin{equation}
\label{Taylor}
D^E_{ij}={1\over 2}\int_0^{\infty}\Biggl[ \langle v_i({\bm x}({\bm a},t),t)\,
v_j({\bm a},0)\rangle
\;+\;\, i\leftrightarrow j\Biggr]\,dt \; .
\end{equation}

Let us now suppose the correlation time $\tau_c$ of the velocity field
to be much smaller than the
sweeping time $\tau_s$. The latter is defined as
\begin{equation}
\label{sweeping}
\tau_s={\lambda\over \langle v^2\rangle^{1/2}}=
\left( {1\over C(0)\int k^2E(k)\,dk}\right)^{1/2}\; ,
\end{equation}
and it
is roughly the average time it takes
for a particle to travel a distance equal
to the correlation length $\lambda$.
The dominant contribution in (\ref{Taylor}) will be given
by Eulerian positions ${\bm x}({\bm a},t)$ close to ${\bm a}$\,:
\begin{equation}
\label{dove}
{\bm x}({\bm a},t)={\bm a}+\int_0^t ds\, {\bm v}({\bm a},s)
+\int_0^t ds\,\left(\nabla_l{\bm v}\right)
({\bm a},s)\int_0^s ds'\,v_l({\bm a},s')
+\ldots\; ,
\end{equation}
and the velocity ${\bm v}({\bm x}({\bm a},t),t)$ in
(\ref{Taylor}) is
\begin{eqnarray}
\label{velocity}
{\bm v}({\bm x}({\bm a},t),t)&=&{\bm v}({\bm a},t) +
\left(\nabla_l{\bm v}\right)
({\bm a},t)\int_0^t ds\,v_l({\bm a},s) + \nonumber \\
&&\left(\nabla_l{\bm v}\right)
({\bm a},t)\int_0^t ds\,\left(\nabla_m v_l\right)({\bm a},s)
\int_0^s ds'\,v_m({\bm a},s') + \nonumber \\
&&{1\over 2}\left(\nabla_l\nabla_m{\bm v}\right)
({\bm a},t)\int_0^t ds\,v_l({\bm a},s)
\int_0^t ds'\,v_m({\bm a},s')+ \ldots \; .
\end{eqnarray}
Eq. (\ref{velocity}) can now be plugged into Taylor's
expression (\ref{Taylor}),
leading to
\begin{eqnarray}
\label{finale}
\!\!\!\!\!D^E_{ij}\!\!\!\!\!&=&\!\!\!\!\!{1\over 2}\int_0^{\infty} dt\Biggl[
\langle v_i({\bm a},t)\,v_j({\bm a},0)\rangle + \nonumber \\
&&\!\!\!\!\!{1\over 2}\int_0^t ds\,\int_0^t ds'\,
\langle(\nabla_l\nabla_m\,v_i)({\bm a},t)\,v_j({\bm a},0)\rangle
\langle v_l({\bm a},s)\,v_m({\bm a},s')\rangle + \nonumber \\
&&\!\!\!\!\!\int_0^t ds\int_0^s ds'
\langle(\nabla_l v_i)({\bm a},t)v_m({\bm a},s')\rangle
\langle(\nabla_m v_l)({\bm a},s)v_j({\bm a},0)\rangle\Biggr]
+ i\leftrightarrow j\, ,
\end{eqnarray}
where homogeneity, incompressibility and
the properties of Gaussian statistics have been exploited.
Eq. (\ref{finale}) is valid for a generic Gaussian random flow
and the next terms in the expansion are $O\left(\tau_c/\tau_s\right)^4$.
Let us now specialize
(\ref{finale}) to the isotropic case. The eddy-diffusivity tensor is
then proportional to $\delta_{ij}$ and its trace
can be calculated by using (\ref{spectrum}) and (\ref{Fourier})\,:
\begin{eqnarray}
\label{isotropo}
{\rm Tr}\,D^E_{ij} &=& 2 \int E(k)\,dk
\int_0^{\infty}dt\,C(t) - \nonumber \\
&-&{2\over 3}\left(\int E(k)\,dk\right)\left(\int k^2E(k)\,dk\right)
\int_0^{\infty}dt\,C(t)\int_0^t ds\,
\int_0^t ds'\,C(|s-s'|)+ \nonumber \\
&+&{1\over 6}\left(\int H(k)\,dk\right)^2\int_0^{\infty}
dt\int_0^t ds\,C(s)\int_0^s ds'
\,C(t-s')\; .
\end{eqnarray}
In order to estimate the order of magnitude of the various terms in
(\ref{isotropo}) it is convenient to
consider the case
\begin{equation}
\label{piatto}
C(|t|)={1\over 2\tau_s}\chi_{\tau_s}(|t|)\,;\quad
\chi_{\tau_s}(|t|)=\cases{1,&if $|t|\leq\tau_s$;\cr
0,&otherwise.\cr}
\end{equation}
When $\tau_s\to 0$ a flow $\delta$-correlated in time is obtained.
In this limit, the only non-vanishing contribution in (\ref{isotropo})
is the first one, which coincides with (\ref{eccoci}).
Both corrections in (\ref{isotropo}) are
proportional
to $\left(\tau_c / \tau_s\right)^2$, as can be checked
by using (\ref{Keith}) and the Schwartz inequality.
If the correlation function $C(t)$ is not
(\ref{piatto}), the constants are changed but not the orders
of magnitude, provided the condition $\int C(t)\,dt=1$ is kept fixed.

Note that for the correlation function (\ref{piatto})
the helicity contribution in (\ref{isotropo})
is clearly positive while the one related to the correlation length
is negative.
These results have a simple physical interpretation.
It is convenient to consider
the Lagrangian correlation time which is proportional to the
eddy-diffusivity.
The first correction in (\ref{isotropo}) is due to the fact
that the presence of a spatial correlation length obviously
reduces the Lagrangian correlation time. The second correction
depends on the presence of helicity. The latter
has the effect that particles
move following a helix, instead of a straight line.
The mean velocity is however the same since it depends
on the energy spectrum only.
It will then take a longer time for a particle to escape a
strongly correlated region, the Lagrangian correlation time is
longer and the eddy-diffusivity is increased.
An equivalent remark is that
a path following a helix is discriminated against tightly bending
back on itself \cite{RHK1}.

A Gaussian flow has been considered but the results
can be easily generalized to the general case. If third-order moments
do not vanish the first correction
will in general be proportional to $\tau_c / \tau_s$.
We finally note that the helical term is actually the only one
in (\ref{finale}) which needed to be symmetrized with respect to
the indices $i$ and
$j$. The latter fact is related
to Onsager's reciprocity theorem \cite{dGM}.
When the helicity does not vanish, the correlation function
will indeed not satisfy the time-reversibility
condition $B_{ij}({\bm x})=B_{ji}({\bm x})$ and the velocity field has
a preferred sense of rotation \cite{M83}.
The lack of parity invariance can be thus
interpreted as a lack of symmetry with respect to time-reversal, which
is responsible for the antisymmetry of transport
coefficients.

\newpage
\centerline{FIGURE CAPTIONS}

\noindent
{\bf FIGURE 1.}  The diagonal component $D^{E}_{11}$ as a function of the bare
molecular
                 diffusivity $D_0$ for the three dimensional ABC flow
(\ref{eq:A.3})
                 with $A=B=C=1$.
                 The continuous line is the result of the Pad\'e method. The
black and white
                 squares are the results of the conjugate gradient method and
                 direct simulations, respectively.

\noindent
{\bf FIGURE 2.}  The diagonal component $D^{E}_{11}$ as a function of the bare
molecular
                 diffusivity $D_0$ for the two dimensional BC flow
(\ref{eq:A.4}) with
                 $B=C=1$.
                 The continuous line is the result of the Pad\'e method.
                 The black squares are the results of the
                 direct simulations.

\noindent
{\bf FIGURE 3.}  The diagonal component $D^{E}_{11}$ as a function of the bare
molecular
                 diffusivity $D_0$ for the two dimensional time dependent flow
(\ref{eq:A.5}).
                 The continuous line is the result of the Pad\'e method.
                 The black and white squares are the results of the conjugate
gradient
                 method and of direct simulations, respectively.

\noindent
{\bf FIGURE 4.} Log-log plot of the energy spectrum of the auxiliary
                field ${\bf w}$ for the (\ref{eq:A.5}) flow at $D_0=0.01$.

\noindent
{\bf FIGURE 5.} The diagonal component $D^{E}_{11}$ as a function of the bare
molecular
                diffusivity $D_0$ for the three dimensional flow
(\ref{eq:zas}),
                (\ref{eq:strezas}) with $\epsilon = 1$ (standard diffusion).

\noindent
{\bf FIGURE 6.}  The diagonal component $D^{E}_{11}$ as a function of the bare
molecular
                 diffusivity $D_0$ for the dimensional flow (\ref{eq:zas}),
                 (\ref{eq:strezas}) with $\epsilon = 2.3$ (anomalous
diffusion).

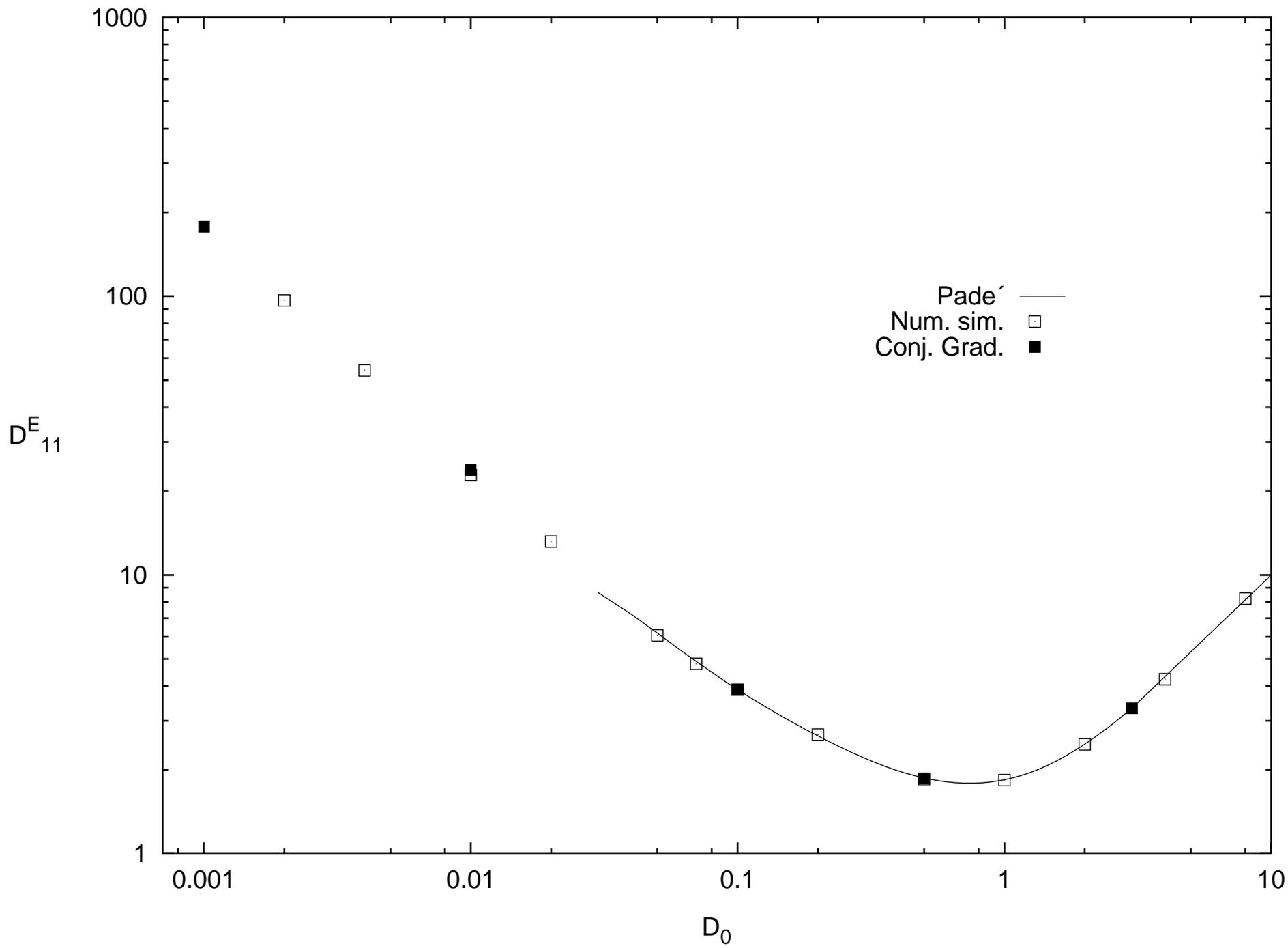
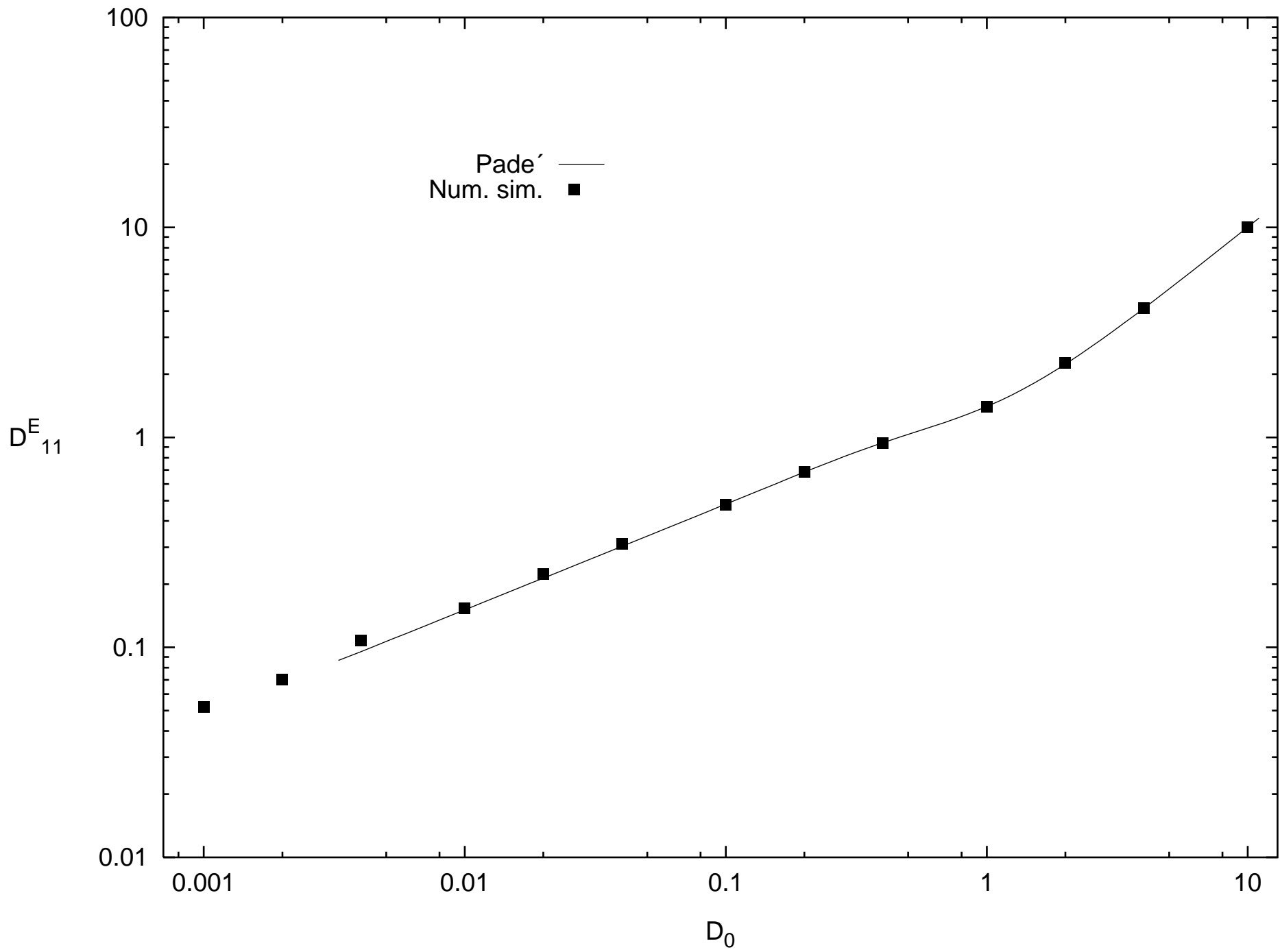
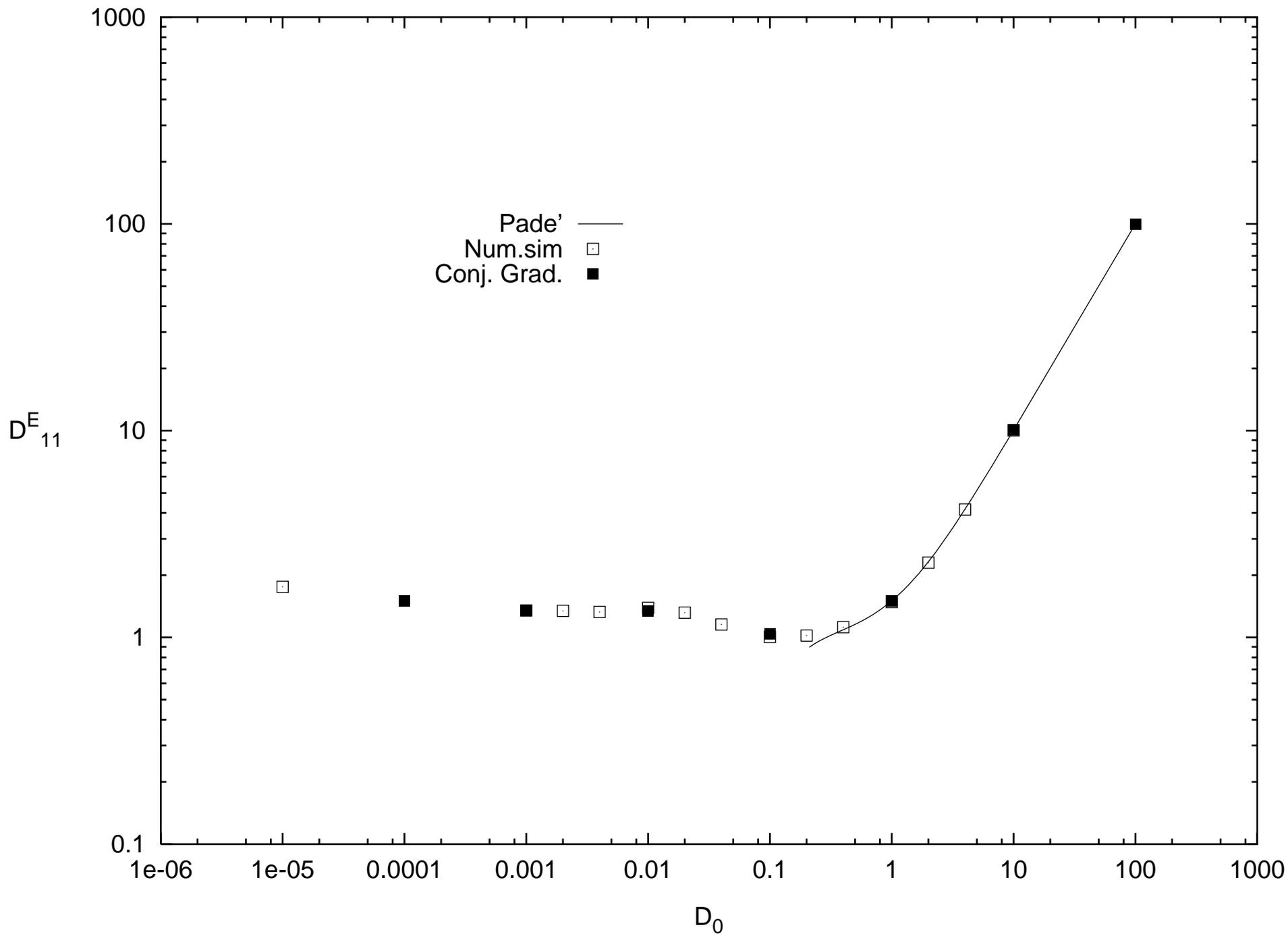
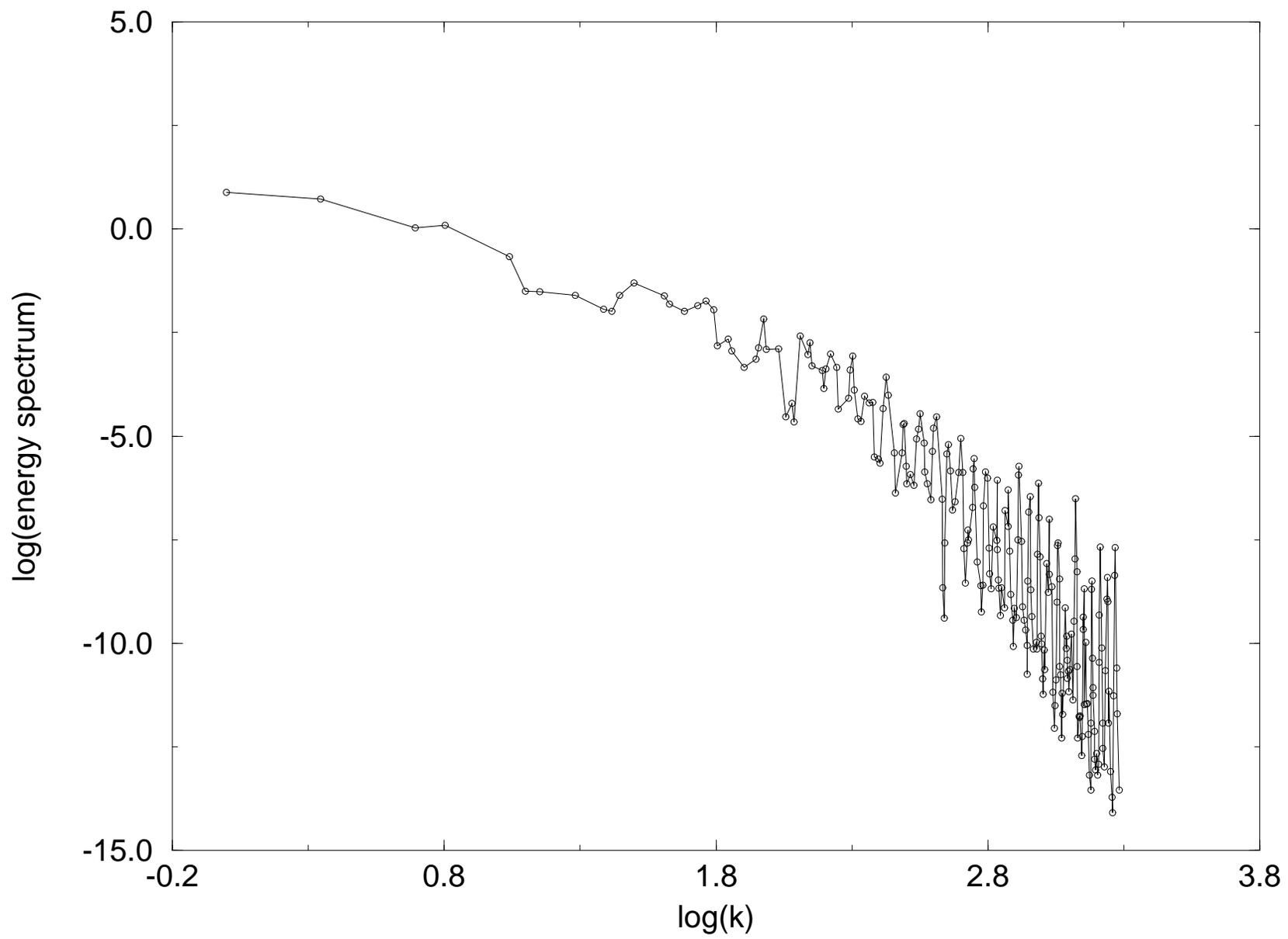
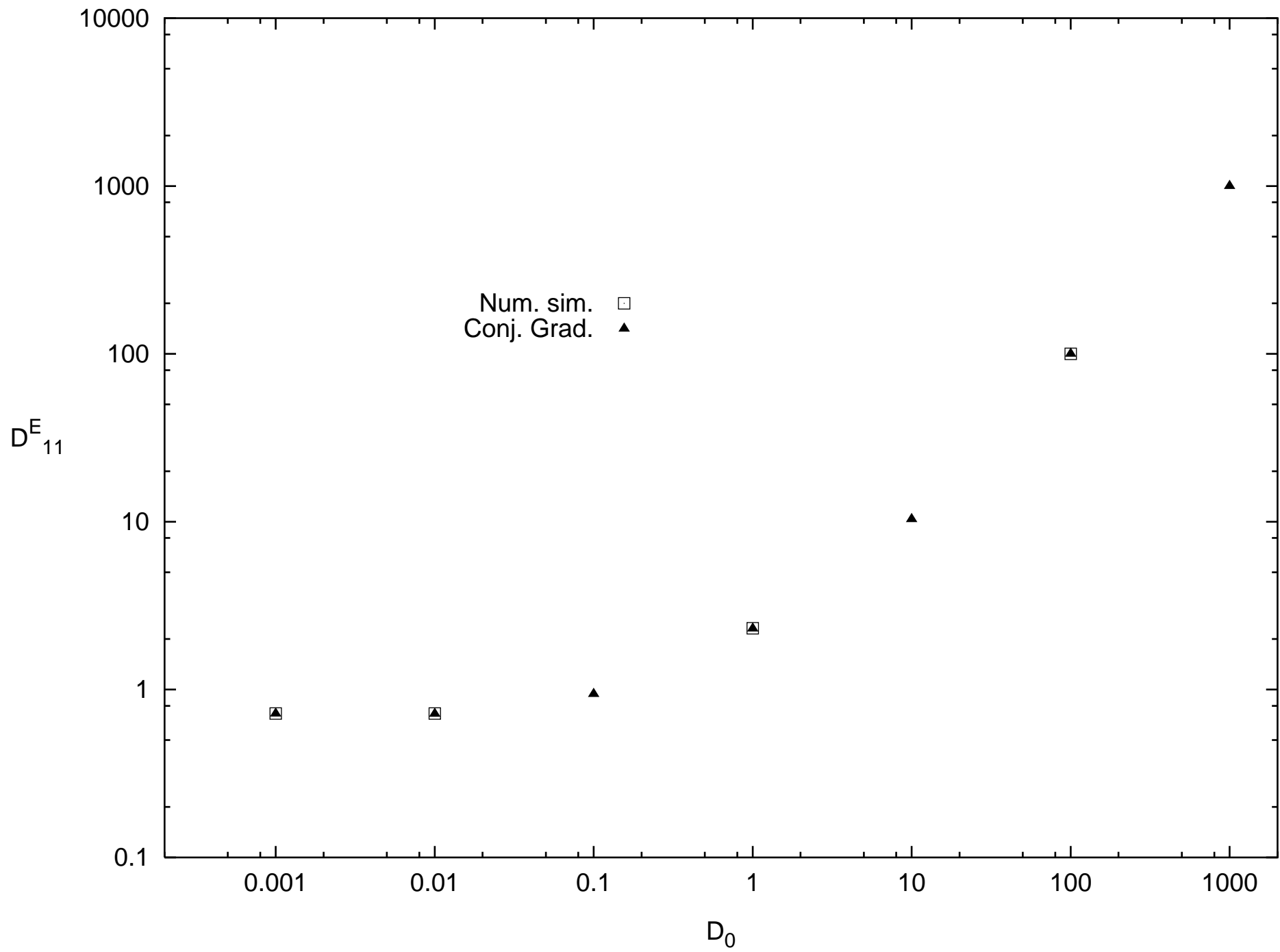
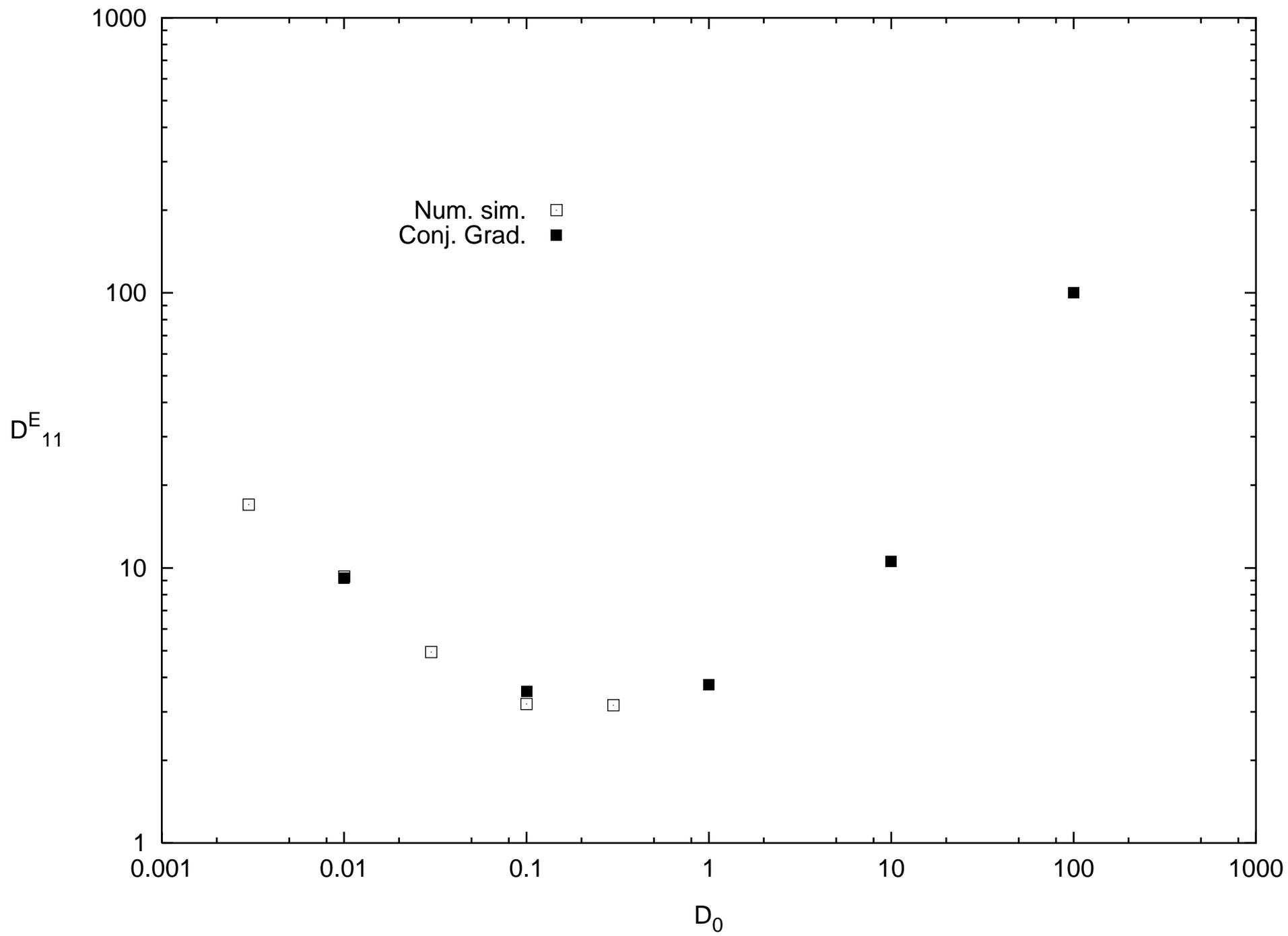
\end{document}